\documentclass[twocolumn,showpacs,amsmath,amssymb,prl]{revtex4}

\usepackage{graphicx}
\usepackage{dcolumn}
\usepackage{bm}

\begin{document}


\title{Crossover from spin accumulation into interface states to spin injection in the germanium conduction band}

\author{A. Jain$^{1}$, J.-C. Rojas-Sanchez$^{1}$, M. Cubukcu$^{1}$, J. Peiro$^{2}$, J. C. Le Breton$^{2}$, E. Prestat$^{1}$, C. Vergnaud$^{1}$, L. Louahadj$^{1}$, C. Portemont$^{3}$, C. Ducruet$^{3}$, V. Baltz$^{4}$, A. Barski$^{1}$, P. Bayle-Guillemaud$^{1}$, L. Vila$^{1}$, J.-P. Attan\'e$^{1}$, E. Augendre$^{5}$, G. Desfonds$^{6}$, S. Gambarelli$^{6}$, H. Jaffr\`es$^{2}$, J.-M. George$^{2}$ and M. Jamet$^{1*}$}
\affiliation{$^{1}$INAC/SP2M, CEA-UJF, F-38054 Grenoble, France\\
$^{2}$UMP CNRS-Thal\`es, CNRS, F-91767 Palaiseau, France\\
$^{3}$CROCUS-Technology, F-38025 Grenoble, France\\
$^{4}$INAC/Spintec, CEA-CNRS-UJF-INPG, F-38054 Grenoble, France\\
$^{5}$LETI, CEA, Minatec Campus, F-38054 Grenoble, France\\
$^{6}$INAC/SCIB, CEA-UJF, F-38054 Grenoble, France}%

\date{\today}

\begin{abstract}
Electrical spin injection into semiconductors paves the way for exploring new phenomena in the area of spin physics and new generations of spintronic devices. However the exact role of interface states in spin injection mechanism from a magnetic tunnel junction into a semiconductor is still under debate. In this letter, we demonstrate a clear transition from spin accumulation into interface states to spin injection in the conduction band of $n$-Ge. We observe spin signal amplification at low temperature due to spin accumulation into interface states followed by a clear transition towards spin injection in the conduction band from 200 K up to room temperature. In this regime, the spin signal is reduced down to a value compatible with spin diffusion model. More interestingly, we demonstrate in this regime a significant modulation of the spin signal by spin pumping generated by ferromagnetic resonance and also by applying a back-gate voltage which are clear manifestations of spin current and accumulation in the germanium conduction band.
\end{abstract}

\maketitle

The first challenging requirement to develop semiconductor (SC) spintronics\cite{Zutic2004,Awschalom2007} \textit{i.e.} using both carrier charge and spin in electronic devices consists in injecting spin polarized electrons in the conduction band of a SC at room temperature. SCs should be further compatible with silicon mainstream technology for implementation in microelectronics making silicon, germanium and their alloys among the best candidates\cite{Zutic2011}. In Si, due to low spin-orbit coupling, very long spin diffusion lengths were predicted and measured experimentally\cite{Dash2009,Jonker2007,Appelbaum2007,Suzuki2011}. Germanium exhibits the same crystal inversion symmetry as Si, low amount of nuclear spins but higher carrier mobility and larger spin-orbit coupling which should allow in principle spin manipulation by electric fields such as the Rashba field\cite{Zhou2011,Jain2011,Saito2011,Jeon2011}. Beyond the conductivity mismatch issue\cite{Fert2001}, another obstacle to electrical spin injection in SCs is the presence of localized states at the interface between the injecting electrode and the SC\cite{Dash2009,Jain2011,Tran2009}. Their exact role in spin injection mechanism and their existence itself need to be elucidated before achieving efficient spin injection in SCs.
In this work, we demonstrate a clear transition from spin accumulation into interface states to spin injection in the conduction band of $n$-Ge. We first observe spin signal amplification at low temperature due to spin accumulation into interface states followed by a clear transition towards spin injection in the conduction band from 200 K up to room temperature. In this regime, the spin signal is reduced down to a value compatible with spin diffusion model. Moreover we demonstrate in this regime a significant modulation of the spin signal by spin pumping generated by ferromagnetic resonance and also by applying a back-gate voltage which are clear manifestations of spin current and accumulation in the conduction band of $n$-Ge. 

\begin{figure}[h!]
\includegraphics[width=0.48\textwidth]{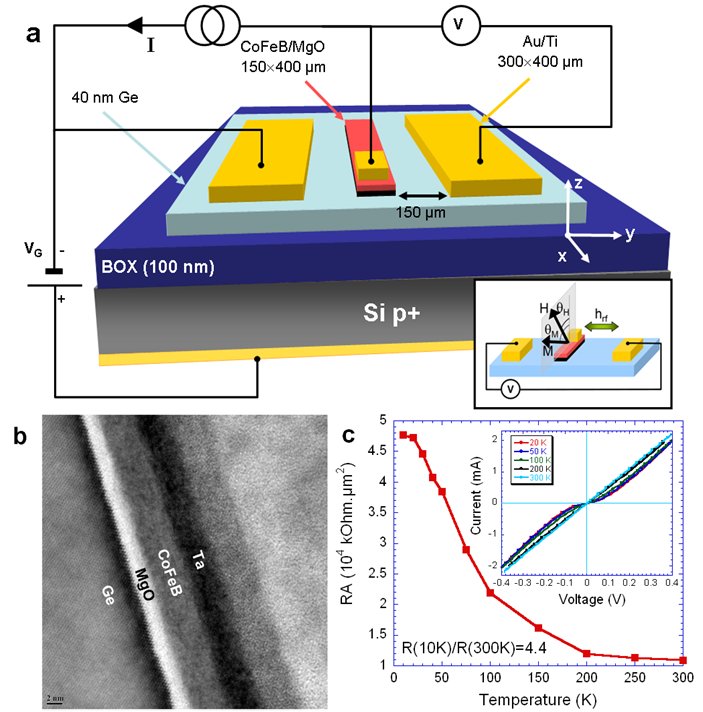} 
\caption{(color online) (a) Schematic drawing of the multi-terminal device used for electrical spin injection, detection and manipulation in germanium. BOX is for buried oxide layer which is made of SiO$_{2}$. The inset shows the geometry for spin pumping measurements. (b) Cross section TEM image of the full stack Ta/CoFeB/MgO/Ge. MgO is poorly crystallized and appears as amorphous in TEM images. (c) Temperature dependence of the magnetic tunnel junction resistance-area $RA$ product measured at a bias current of 50 $\mu$A. Inset: $I(V)$ curves recorded at different temperatures between the tunnel junction and an ohmic contact.}\label{fig1}
\end{figure}

The multi-terminal device we used for electrical measurements is shown in Fig. 1a. The full stack Ta(5 nm)/Co$_{60}$Fe$_{20}$B$_{20}$(5 nm)/MgO(3 nm) has been grown by sputtering and annealed on Germanium-On-Insulator (GOI) wafers\cite{Jain2011,note1} (Fig. 1b). We have inserted a thin MgO tunnel barrier to circumvent the conductivity mismatch and partly alleviate the Fermi level pinning by reducing the interface states density\cite{Cantoni2011,Lee2010,Zhou2010} which leads to a modest Schottky barrier height at the MgO/$n$-Ge interface. Conventional optical lithography was used to define 3-terminal devices made of a tunnel spin injector in between two ohmic contacts made of Au(250 nm)/Ti(10 nm). Fig. 1c displays the tunnel junction $RA$ product and the corresponding $I(V)$ curves at different temperatures. We first observe clear non-linear symmetric $I(V)$ characteristics which confirms that tunnelling transport takes place in our junctions. Furthermore the $RA$ product only varies by a factor 4.4 between 10 K and 300 K again in good agreement with tunnelling dominated transport. It exhibits a transition to low $RA$ values above 200 K. The 30 nm-thick Ge channel exhibits a metallic character with its resistivity increasing by 18.7 \% with temperature and reaching 3.7 m$\Omega$.cm at room temperature. The associated electron mobility is of the order of 1700 cm$^{2}$.V$^{-1}$.s$^{-1}$.

\begin{figure}[h!]
\includegraphics[width=0.48\textwidth]{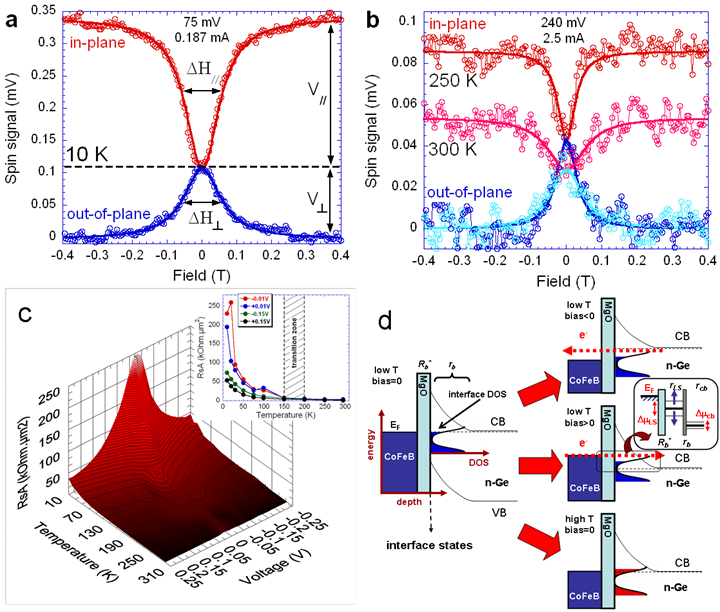} 
\caption{(color online) (a) and (b) Hanle and inverted Hanle curves recorded at 10 K and 250 K, 300 K respectively. Solid lines are Lorentz fits. $\Delta$H$_{\bot}$ and $\Delta$H$_{//}$ are the Full Width at Half Maximum (HWHM) of Hanle and inverted Hanle curves. (c) Spin resistance-area product as a function of temperature and applied voltage. This 3D plot is the result of more than 5.10$^{5}$ measurements on the same device. The inset shows the temperature dependence of $R_{S}A$ at 4 different bias voltages. (d) Sketch of the energy band diagram of CoFeB/MgO/Ge and its modification upon bias voltage and increasing temperature. Following Ref. \cite{Matsubara2008} and \cite{Taoka2011}, the interface density of states (DOS) has a characteristic U-shape with a minimum at the Ge mid-gap. At low T and for positive bias voltage, the detailed resistor model is given at the CoFeB/MgO/Ge interface showing spin accumulations into interface states and in the Ge conduction band.} \label{fig2}
\end{figure}

Electrical spin injection/detection measurements have been performed at different temperatures and bias voltages using the contacts geometry displayed in Fig. 1a. Spin detection is achieved by Hanle measurements that probe spin accumulation at the ferromagnet/SC interface. In comparison, inverse spin Hall effect (ISHE) and gate effect presented in the second part of this letter are sensitive to spin accumulation far from the interface in the germanium conduction band. The 3-terminal geometry in which the same electrode is used for spin injection and detection, represents a unique tool to probe spin accumulation both into interface states and in the channel. In Fig. 2a and 2b, the magnetic field was applied out-of-plane along $z$ to obtain Hanle curves ($V_{\bot}$: spin precession around the applied field) and in-plane along $x$ to obtain inverted-Hanle curves ($V_{//}$: to suppress spin precession around interface random fields)\cite{Dash2011}. In that case, the total spin signal scaling with the full spin accumulation is given by: $V_{S}=V_{//}+V_{\bot}$ and the spin resistance-area product by: $R_{S}A=(V_{S}/I).A$ where $I$ is the applied current and $A$ the ferromagnetic electrode area. All our measurements of the total spin signal have been gathered into a single 3D plot in Fig. 2c that clearly demonstrates spin signal amplification at low bias and below 150 K. This spin amplification is usually attributed to sequential tunnelling through localized states at the MgO/Ge interface\cite{Dash2009,Jain2011,Tran2009} due to spin confinement by the reminiscent Schottky barrier. Using the same spin diffusion model as in Ref. \cite{Tran2009}, spin accumulations into localized states and in the Ge conduction band are respectively given by (see Fig. 2d and Ref. \cite{note2}):

\begin{eqnarray*}
& \Delta\mu_{LS}\approx 2e\times TSP\times j\frac{r_{LS}(r_{cb}+r_{b})}{r_{LS}+r_{cb}+r_{b}} \\
& \Delta\mu_{cb}\approx 2e\times TSP\times j\frac{r_{LS}r_{cb}}{r_{LS}+r_{cb}+r_{b}}
\end{eqnarray*}

It should be underlined here that these expressions are only valid in the case of a two-step tunneling process through interface states. $TSP\approx$0.65 is the tunnel spin polarization estimated from the Julli\`ere's formula\cite{Julliere1975} on a symmetric magnetic tunnel junction CoFeB/MgO/CoFeB at room temperature and $j$ is the current density. The spin accumulation $\Delta\mu$ is related to the experimental spin signal $V_{S}$ through the expression: $V_{S}=TSP\times\Delta\mu/2e$ (Ref. \cite{Dash2009}). Both spin amplification and spin injection in the conduction band depend on the relative intensity of $r_{LS}$, $r_{b}$ and $r_{cb}$ as well as on their temperature and bias voltage dependence. As shown in Fig. 2d, at low temperature and low bias voltage, the two-dimensional density of states (2D-DOS) of localized states at the Fermi level is low ($\mathcal{N}^{LS}\ll\mathcal{N}^{cb}$) so that $r_{LS}$, $r_{b}\gg r_{cb}$ and the spin-current leakage from the localized states to the conduction band through the Schottky barrier is very weak ($\tau_{\rightarrow}^{LS}\gg\tau_{sf}^{LS}$) leading to $r_{b}\gg r_{LS}$. Hence we find: $\Delta\mu=\Delta\mu_{LS}\approx 2e\times TSP\times jr_{LS}$ (spin signal amplification due to spin accumulation into interface states) and $\Delta\mu_{cb}\approx 2e\times TSP\times j(r_{cb}/r_{b})r_{LS}\ll \Delta\mu_{LS}$ (spin accumulation in the Ge conduction band is negligibly small and only spin accumulation into interface states is detected). Still at low temperature but now at finite bias voltage as shown in Fig. 2d, sequential tunnelling takes place through much higher interface 2D-DOS leading to a drastic reduction of $r_{LS}$ and $\Delta\mu =\Delta\mu_{LS}$. In the same way, by increasing the temperature, electrons may tunnel through higher energy levels at the interface where the 2D-DOS is much higher which also reduces $r_{LS}$ and $\Delta\mu =\Delta\mu_{LS}$ (see Fig. 2d). We then infer that the spin signal transition we observe between 150 K and 200 K is due to the progressive decrease of $r_{b}$ with temperature (due to thermally activated electrical transport over the Schottky barrier) as pointed out in ref. \cite{Jain2011} and experimentally observed in the temperature dependence of the $RA$ product (Fig. 1c). Assuming $r_{LS}\gg r_{cb}$ in the whole temperature range which is quite fair regarding the difference in 2D-DOS\cite{Matsubara2008} ($\mathcal{N}^{LS}\approx$ 10$^{11}$-10$^{12}$ cm$^{-2}$.eV$^{-1}\ll\mathcal{N}^{cb}\approx$10$^{14}$ cm$^{-2}$.eV$^{-1}$) we propose the following scenario: \textit{i)} for T$<<$150 K, $r_{b}\gg r_{LS}$, $\Delta\mu=\Delta\mu_{LS}\approx 2e\times TSP\times jr_{LS}$ and $\Delta\mu_{cb}\approx$0 as discussed previously, \textit{ii)} for 150 K$<$T$<$200 K, $r_{b}\approx r_{LS}$, $\Delta\mu_{LS}\approx 2e\times TSP\times jr_{LS}r_{b}/(r_{LS}+r_{b})$ and $\Delta\mu_{cb}\approx 2e\times TSP\times jr_{LS}r_{cb}/(r_{LS}+r_{b})$: we still observe spin signal amplification but spins start to accumulate in the Ge conduction band partly due to a shorter carrier tunnelling transfer time from localized states into the $n$-Ge conduction band and \textit{iii)} for T$>>$200 K, $r_{b}\ll r_{cb}$, $\Delta\mu_{LS}=\Delta\mu_{cb}\approx 2e\times TSP\times jr_{cb}$, interface states progressively couple to the Ge conduction band and a significant spin accumulation now takes place in the channel. The observation of inverse spin Hall effect (ISHE) by spin pumping together with the observation of spin signal modulation upon the application of a gate voltage for temperatures higher than 200 K clearly support this scenario. 

\begin{figure}[h!]
\includegraphics[width=0.48\textwidth]{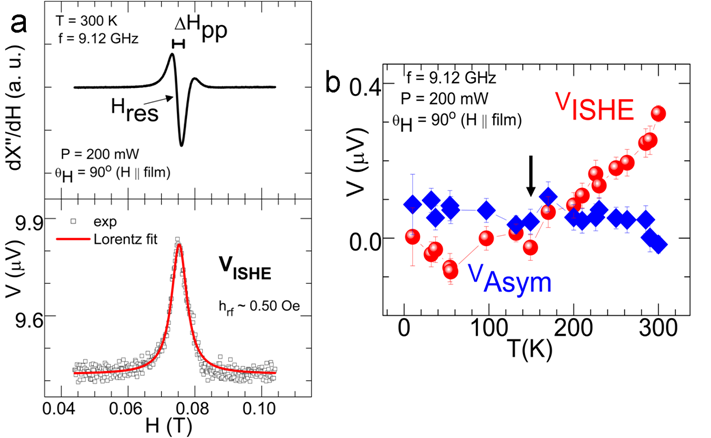} 
\caption{(color online) (a) FMR line and $V_{ISHE}$ measured at room temperature. The applied field is along $x$ ($\theta_{H}$=90$^{\circ}$). $H_{res}$ and $\Delta H_{pp}$ are the resonance field and FMR linewidth respectively. (b) Temperature dependences of the asymmetric voltage $V_{Asym}$ and $V_{ISHE}$.} \label{fig3}
\end{figure}

As shown in Fig. 3a, by measuring simultaneously the voltage between the Au/Ti ohmic contacts and the ferromagnetic resonance (FMR) spectrum of Ta/CoFeB/MgO/Ge we are able to investigate spin injection in Ge via the inverse spin Hall effect (ISHE) at zero bias voltage\cite{Saitoh2006,Vila2007,Ando2011a,Ando2011b}. For this purpose, the same device as the one previously used for electrical measurements is introduced into a Br$\ddot{u}$ker X-band cavity and the measuring geometry is shown in the inset of Fig. 1a. All the measurements presented in Fig. 3 have been performed with the static magnetic field applied along $x$ ($\theta_{H}$=90$^{\circ}$) \textit{i.e.} along the CoFeB bar. Under radiofrequency excitation the magnetization precession in the ferromagnetic layer pumps spins to the non-magnetic Ge layer and the corresponding spin current generates an electric field in Ge due to ISHE: $\bf{E_{ISHE}}\propto\bf{J_{S}}\times\bf{\sigma}$ where $\bf{J_{S}}$ is the spin-current density along $z$ and $\bf{\sigma}$ its spin polarization vector. This electric field $\bf{E_{ISHE}}$ converts into a voltage $V_{ISHE}$ between both ends of the Ge channel. The microwave frequency was $f$=9.12 GHz and we could observe the voltage signal due to ISHE at room temperature (Fig. 3a) and easily fit it using a symmetrical Lorentzian curve. This symmetric voltage coincides with the main FMR line at $H_{res}$=0.074 T\cite{note3}. This result clearly demonstrates the presence of both spin accumulation and related spin current in the Ge conduction band at room temperature. Furthermore we have shown that all our findings are in good agreement with the observation of ISHE: symmetrical behaviour of $V_{ISHE}$ around the resonance field $H_{res}$, $V_{ISHE}$=0 when the external magnetic field is applied perpendicular to the film ($\theta_{H}$=0), $V_{ISHE}$ changes its sign when crossing $\theta_{H}$=0 and finally the linear dependence of its amplitude with the microwave power excitation\cite{Rojas2012}. From $V_{ISHE}$ at room temperature, we adapted a model based on the spin mixing conductance formalism\cite{Tserkovnyak2005} already used in metals\cite{Mosendz2010} and semiconductors with a Schottky contact\cite{Ando2011b,Ando2012} to estimate the spin Hall angle in $n$-Ge: $\theta_{SHE}\approx$0.002 (see Supplemental Material). We found similar spin Hall angles on several devices. This model is probably not the most appropriate and should be reconsidered in the case of Ferromagnet/Oxide/SC systems but it gives a reasonable value for $\theta_{SHE}$ in Ge. This is indeed of the same order of magnitude as in $n$-GaAs (0.007 in Ref. \cite{Ando2011b}) and one order of magnitude larger than in $p$-Si (0.0001 in Ref. \cite{Ando2012}).
We have then investigated the temperature dependence of ISHE. For this purpose, the voltage corresponding to the main FMR peak has been fitted using both symmetric and asymmetric contributions. They are reported as a function of temperature in Fig. 3b. The symmetric signal is related to ISHE while the asymmetric one may be due to reminiscent rectification effects as a combination of radiofrequency eddy currents and anisotropic magnetoresistance (AMR) in CoFeB\cite{Mosendz2010b,Azevedo2011}. Increasing the temperature, we clearly see a transition with the appearance of ISHE at approximately 150 K which is in perfect agreement with the transition from spin accumulation into interface states to the $n$-Ge conduction band discussed previously. The asymmetric voltage contribution remains negligible and almost constant within error bars in the whole temperature range.

\begin{figure}[h!]
\includegraphics[width=0.48\textwidth]{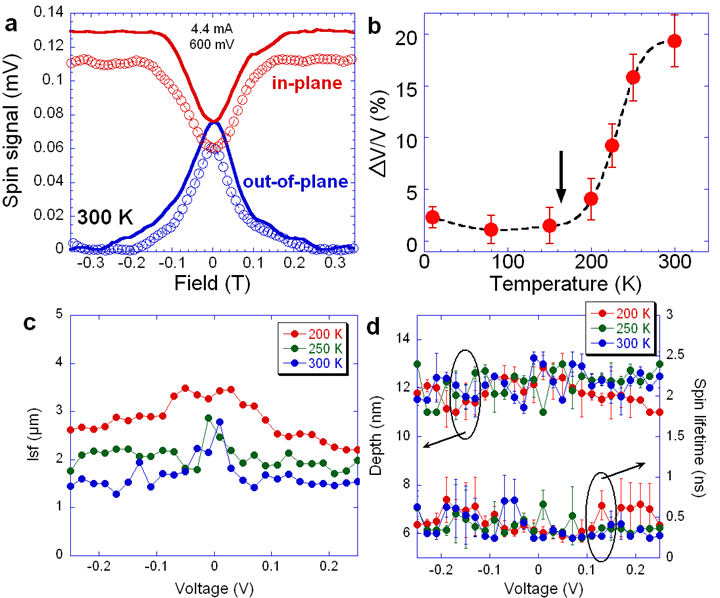} 
\caption{(color online) (a) Effect of a back gate voltage (0 V and -50 V) on Hanle and inverted Hanle spin signals at 300 K. Open symbols are for $V_{G}$=0 V and solid lines for $V_{G}$=-50 V. (b) Temperature dependence of $\Delta V/V$=($V_{S}$(-50V)-$V_{S}$(0V))/$V_{S}$(0V) in \%. The dashed is a guide to the eyes. (c) Spin resistance-area product converted into spin diffusion length $l_{sf}$ using the diffusive model of Ref. \cite{Fert2001} developed for spin injection in the germanium channel. Only T=200 K, 250 K and 300 K values are reported when spin polarized electrons are injected in the Ge conduction band. (d) Results of the triple fit over $r$=$V_{\bot}$/$V_{S}$, $\Delta H_{\bot}$ and $\Delta H_{//}$ yielding the mean depth at which spin polarized electrons are injected and their spin lifetime. Depths and spin lifetimes are almost identical at 200 K, 250 K and 300 K.} \label{fig4}
\end{figure}

Using the field-effect transistor structure, we now focus on the application of a gate voltage to the Ge channel to modulate the spin signal\cite{Ando2011c}. At negative gate voltage to a maximum of $V_{G}$=-50 V, the carrier density is lowered in the $n$-Ge channel and its resistivity is enhanced\cite{note4}. At 10 K, we find:  $\Delta R$/$R$=$R$(-50V)-$R$(0V)/$R$(0V)=+68.2 \% whereas $\Delta R$/$R$=+21.9 \% at 300 K.  The resulting spin signal variation at 300 K is reported in Fig. 4a. We can clearly see the effect of the gate voltage with a significant spin signal increase whereas almost no variation is observed at 10 K (not shown). All the measurements are summarized in Fig. 4b as a function of temperature: a clear transition occurs again between 150 K and 200 K (171 K from a linear fit to the finite values of $\Delta V$/$V$ above 200 K). Again these findings are in good agreement with a transition from spin injection into interface states to the Ge conduction band. To be more quantitative, in the case of spin injection in the Ge conduction band and in the frame of the diffusive regime model\cite{Fert2001}, the spin resistance-area product is given by: $R_{S}A=V_{S}/I.A=(TSP\times l_{sf}^{cb})^{2}(\rho/t_{Ge})$. Hence if we assume that $TSP$ and $l_{sf}^{cb}$ remain constant under the application of an electric field, $V_{S}$ scales as $(\rho/t_{Ge})$ which is proportional to the channel resistance $R$. We thus expect $\Delta V$/$V$ to scale with $\Delta R$/$R$ in the event that spin polarized carriers are injected in the Ge conduction band. Below 150 K, we obtain negligible values of $\Delta V$/$V$ for large values of $\Delta R$/$R$ which is compatible with spin accumulation into interface states. Above 200 K, $\Delta V$/$V$ starts to increase as spins start to accumulate in the Ge conduction band and at room temperature we find $\Delta V$/$V\approx\Delta R$/$R$ which means that we fully achieved spin injection in the Ge conduction band.\\
Based on the results we obtained in spin pumping and electric field effect measurements, we can now estimate $l_{sf}^{cb}$ from the experimental spin resistance-area product as well as $\tau_{sf}^{cb}$ from the fit of Hanle and inverted Hanle curves. In Fig. 4c, we have converted the spin resistance-area product into $l_{sf}^{cb}$ using the spin diffusion model\cite{Fert2001}: $R_{S}A=(TSP\times l_{sf}^{cb})^{2}(\rho/t_{Ge})$ at T=200 K, 250 K and 300 K. At room temperature, we find $l_{sf}^{cb}\approx$ 1.5 $\mu$m with weak bias voltage dependence up to $\pm$0.5 V. Moreover in Hanle and inverted Hanle measurements, spins that are injected electrically precess around the external magnetic field (either along $z$ or $x$) and random fields created by surface charges. Random fields are calculated from the surface roughness parameters given by atomic force microscopy (AFM) performed on the whole stack Ta/CoFeB/MgO/Ge. We found a root-mean-square (RMS) roughness of 0.2 nm and a correlation length of 40 nm. Then the spin dynamics has been computed by considering only spin precession and relaxation: spin drift and diffusion were neglected as discussed in Ref. \cite{Dash2011}. Finally a triple fit over $r$=$V_{\bot}$/$V_{S}$, $\Delta H_{\bot}$ and $\Delta H_{//}$  yields the mean depth at which spin polarized electrons are injected and $\tau_{sf}^{cb}$. The results are displayed in Fig. 4d and are almost the same for all three temperatures: 200 K, 250 K and 300 K with very weak bias voltage dependence. In average, spin polarized electrons are injected 10 nm deep into the Ge film and $\tau_{sf}^{cb}$=400$\pm$100 ps at room temperature. Using $D$=$\mu k_{B}T/e\approx$43 cm$^{2}$.s$^{-1}$, we find: $l_{sf}^{cb}$=1.3$\pm$0.2 $\mu$m at room temperature which is in very good agreement with the value obtained from the spin resistance-area product.\\
In summary, we have demonstrated a clear crossover from spin accumulation into interface states that leads to spin signal amplification to spin injection in the Ge conduction band at 200 K. For that purpose, we have shown inverse spin Hall effect in Ge and spin signal modulation applying a back gate voltage from 200 K up to room temperature. From a general point of view, we believe that the same transition should be observable in any low gap semiconductors in the presence of interface states.\\ 
This work was supported by the Nanoscience Foundation of Grenoble through the RTRA project IMAGE. The initial GOI substrates were obtained through the collaboration with Soitec under the public funded NanoSmart program (French OSEO).


\begin{thebibliography}{}
\bibitem{note2}
$R_{b}^{*}$ and $r_{b}$=$\tau_{\rightarrow}^{LS}/(e^{2}\mathcal{N}^{LS})$ are the tunnel and Schottky resistances respectively. $\Delta\mu_{LS}=\mu_{LS}^{\uparrow}-\mu_{LS}^{\downarrow}$ and $r_{LS}=\tau_{sf}^{LS}/(e^{2}\mathcal{N}^{LS})$ (resp. $\Delta\mu_{cb}=\mu_{cb}^{\uparrow}-\mu_{cb}^{\downarrow}$ and $r_{cb}=\rho l_{sf}^{cb}=\tau_{sf}^{cb}/(e^{2}\mathcal{N}^{cb})$) are the spin accumulation and spin resistance of localized interface states (resp. Ge conduction band). $\tau_{\rightarrow}^{LS}$ is the characteristic leakage time from localized states to the Ge conduction band and depends on the Schottky barrier height and width, $\mathcal{N}^{LS}$ is the two-dimensional density of states (2D-DOS) at the MgO/Ge interface, $\tau_{sf}^{LS}$ is the spin lifetime of electrons into localized states, $\rho$ the channel resistivity and $l_{sf}^{cb}$ (resp. $\tau_{sf}^{cb}$) the spin diffusion length (resp. spin lifetime) of electrons in $n$-Ge. $\mathcal{N}^{cb}$ is the 2D-DOS in the Ge conduction band.
\bibitem{note3}
A second but very weak FMR line appears in some devices as the one of Fig. 3a at $H_{res}$=0.077 T. This resonance peak systematically gives a weak asymmetric voltage contribution that decreases with temperature. The origin of this secondary peak is not investigated in this paper.
\bibitem{note4}
The field-effect transistor exhibits non-symmetric behavior and at positive gate voltage (up to $V_{G}$=50 V) the channel resistivity remains almost unchanged. We have also performed non-local $I(V)$ measurements to probe the interface resistance under the application of a gate voltage. For positive and negative gate voltage, we found the same interface resistance.
\end{thebibliography}
\end{document}